\documentclass[10pt,leqno]{amsart}

\usepackage{graphicx}
\usepackage{indentfirst,csquotes}
\usepackage{fancyhdr}
\usepackage[utf8]{inputenc}
\usepackage{natbib}
\usepackage{hyperref}
\usepackage{amsmath,amssymb,amsthm}
\usepackage{autobreak}
\usepackage{amsfonts}
\usepackage{booktabs}
\usepackage{color}
\usepackage[table]{xcolor}
\usepackage{multirow}
\usepackage{subcaption}
\usepackage[export]{adjustbox}
\usepackage{setspace}

\pdfoutput=1

\hypersetup{
    colorlinks=true,
    linkcolor=black,
    filecolor=black,
    urlcolor=black
}

\def\bSig\mathbf{\Sigma}

\newcommand{\bfs}{\hbox{\boldmath$s$}}

\newcommand{\bfR}{\hbox{\boldmath$R$}}

\newcommand{\bfQ}{\mathbf{Q}}

\newcommand{\bfw}{\hbox{\boldmath$w$}}
\newcommand{\bfz}{\hbox{\boldmath$z$}}
\newcommand{\bfM}{\hbox{\boldmath$M$}}

\newcommand{\bfG}{\mathbf{G}}

\newcommand{\bfzero}{\hbox{\boldmath$0$}}
\newcommand{\bfpi}{\hbox{\boldmath$\pi$}}

\newcommand{\bfphi}{\hbox{\boldmath$\bfphi$}}

\newcommand{\bftheta}{\hbox{\boldmath$\theta$}}

\newcommand{\bfLambda}{\hbox{\boldmath$\Lambda$}}

\title{Incorporating animal movement into continuous-time spatial capture-recapture models}

\author{
Clara Panchaud$^{1,2,*}$,
Ruth King$^{2}$,
David Borchers$^{3,4}$,
and Hannah Worthington$^{3,4}$
\vspace{10pt}
\\
{\small
$^{1}$Department of Statistics, University of British Columbia, Vancouver, British Columbia V6T 1Z4 Canada \\
$^{2}$School of Mathematics and Maxwell Institute, University of Edinburgh, Edinburgh EH9 3FD, Scotland \\
$^{3}$School of Mathematics and Statistics, University of St Andrews, Fife, KY16 9LZ, Scotland \\
$^{4}$Centre for Research into Ecological and Environmental Modelling, University of St Andrews, Fife, KY16 9LZ, Scotland \\
$^{*}$Corresponding author: \texttt{clara.panchaud@stat.ubc.ca}
}
}

\date{}

\begin{document}

\begin{abstract}

Estimation of wildlife population size and spatial dynamics is central to ecology and conservation. Spatial capture–recapture (SCR) models estimate abundance, using detections at sensors such as camera traps, by linking detection probability to the distance between detectors and latent individual activity centres. However, standard SCR formulations assume detections are conditionally independent over time given activity centres without explicitly modeling movement between detection events. This assumption can be problematic when individuals exhibit movement-driven dependence in detections, potentially leading to biased inference on population size. We address this important issue by developing a continuous-time framework that integrates movement into spatial capture–recapture. Individual movement is modeled as a continuous-time Markov chain over a discretised landscape, and detections arise as state-dependent Poisson events. This yields a Markov-modulated marked Poisson process representation, in which detections provide information about an individual’s latent location at the time of observation and allow likelihood-based inference in continuous time. We show through simulation studies that ignoring movement-driven dependence can lead to positively biased estimates of population size, whereas the proposed model recovers unbiased estimates and provides additional inference on space use. An application to camera-trap data of American martens illustrates how the framework yields new insights into movement and density. These results demonstrate that explicitly modeling movement is critical for reliable inference in spatial capture–recapture studies.

\end{abstract}


\maketitle``

\section{Introduction}
\label{s:introduction}

Quantifying animal abundance and spatial dynamics is central to ecological inference and conservation planning. Capture–recapture (CR) models provide a statistical framework for estimating population size from repeated detections of identifiable individuals while accounting for imperfect detection \citep{seber1982estimation, chao2001overview, mccrea2014analysis}. When detection locations are recorded, spatial capture–recapture (SCR) models allow estimation of population density by linking detection probability to the distance from an individual’s latent activity centre \citep{efford2004density, borchers2008spatially, royle2013spatial}.

Although SCR has been highly successful across a wide range of species and monitoring designs, its standard formulation assumes a static latent activity centre and detections that are conditionally independent over time given this centre \citep{efford2004density, borchers2008spatially, royle2013spatial}. This independence assumption implies that the detection process contains no residual temporal dependence once the activity centre is accounted for. In practice, this assumption is often violated because individuals move through their environment, inducing dependence in their locations over time and space. As a result, detections close in time are more likely to arise from nearby locations, generating movement-driven correlation that is not captured by static activity centre models. Ignoring this structure can affect inference and lead to biased parameter estimates \citep{moqanaki2021consequences, stevenson2021spatial, panchaud2024incorporating}. Addressing this issue requires models that explicitly account for the movement processes generating such dependence, allowing temporal and spatial correlation in detections to be modeled directly. We adopt a likelihood-based framework which enables joint inference on population size and movement.

While SCR models can be used to predict the locations of latent activity centres and to quantify the space use of individuals over the survey period, they do not explicitly describe how individuals move through the landscape over time. Consequently, the movement processes underlying detections are not directly modeled. Incorporating animal movement models into SCR requires linking the detection process to a time-evolving latent location, which is only partially observed through detections at sensors. These limitations of standard SCR models have motivated recent work on integrating movement processes into SCR.

SCR models that incorporate animal movement models have received growing attention in recent years, notably highlighted in the Special Feature “Linking capture–recapture and movement” in Ecology \citep{converse2022special}. \citet{mcclintock2022integrated} outlined a general framework for integrating movement and SCR models, while \citet{dupont2022improved}, \citet{hostetter2022modeling}, and \citet{chandler2022modeling} developed approaches that incorporate telemetry data into SCR analyses. \citet{gardner2022integrated} modeled movement by allowing the activity center to vary over time, and discrete-time state–space models have also been applied in this context \citep{crum2023forecasting}. Other approaches have relaxed the conditional independence assumption without explicitly modeling movement, for example using Hawkes processes \citep{van2026spatial} or Ornstein–Uhlenbeck processes \citep{panchaud2024incorporating}. Although these developments represent important advances, a fully continuous-time likelihood-based SCR framework that directly represents individual movement without relying on external telemetry data has yet to be developed. In this paper, we introduce a continuous-time state–space SCR framework, referred to as model \emph{Move–SCR}, that integrates movement into SCR using a Markov-modulated marked Poisson process.

The proposed framework builds on continuous-time state–space modeling, a natural framework for representing latent ecological processes \citep{newman2014modelling, auger2021guide, newman2023state}. State–space models describe the evolution of an unobserved state process together with an observation process conditional on the latent state, and have been widely applied to a range of ecological types of data \citep{de2002fitting, royle2008modeling, patterson2008state, king2012review, king2014statistical, wilson2018estimating}. A particularly flexible continuous-time state–space model is the Markov-modulated Poisson process (MMPP), in which event rates depend on the current state of a latent Markov process \citep{neuts1979versatile, fischer1993markov}. MMPPs are well suited for time-to-event data and have been applied in diverse settings such as whale surfacing \citep{langrock2013markov} and earthquake occurences \citep{lu2012markov}. An extension of this framework is the Markov-modulated marked Poisson process (MMMPP) \citep{freed1982poisson}, which associates each event with a mark whose distribution depends on the current latent state. MMMPPs have been used in genetics, disease modeling, and continuous-time capture–recapture contexts \citep{fearnhead2006exact, choquet2018markov, mews2023markov}.

The proposed \emph{Move–SCR} framework utilises the MMMPP framework to describe animal movement within the context of camera trap data. The method is likelihood-based and naturally accommodates irregular detection times while explicitly representing the underlying movement process. The latent process describes an individual animal's movement as a continuous-time Markov chain over a discretised landscape, where each state corresponds to a discretised spatial grid cell. The model converges to a movement model on continuous-space as the state space size tends to infinity. Transitions among cells follow the specified underlying movement process, and detections arise as Poisson events conditional on the individual occupying a cell containing a camera trap. The mark associated with each event is the known trap location, so that each detection provides information about the latent spatial state at the time of observation.

The remainder of this paper is organized as follows. Section~\ref{s:methods} presents the novel \emph{Move–SCR} model within the state–space framework, including the transition and observation processes. Section~\ref{s:model fitting} describes likelihood-based estimation and computational implementation. Simulation studies are presented in Section~\ref{MMMPP:simulations}, followed by an application to camera-trap data of American martens in Section~\ref{MMMPP:application}. Section~\ref{MMMPP:discussion} concludes with a discussion.

\section{Methods}
\label{s:methods}

\subsection{Notation}
\label{s:notations}

A camera trap survey occurs over a region $\mathcal{R}$ of area $A$, with $K$ camera traps placed at fixed locations $\{\bfz_k, \ k=1\dots,K\}$. We assume all camera traps remain active over the survey duration $[t_0=0, T]$. The parameter of primary interest is the unknown total population size $N$. 
Each individual $i\in\{1,\dots,N\}$ is associated with a latent activity centre $ \bfs_i \in \mathcal{R}$.

The observed data consist of the capture histories of $n \le N$ observed individuals, comprising exact times and associated camera trap location. The number of detections of an individual $i\in \{1,\dots, n \}$ is denoted by $J_i$. One detection is represented as a vector $\bfw_{ij} = (t_{ij}, x_{ij}), j\in \{ 1, \dots, J_i\}$, where $t_{ij}$ is the time of the detection and $x_{ij}\in \{1,\dots, K\}$ the index of the trap at which the detection occurred. We note that we use the terms detection and observation interchangeably throughout. The capture history of individual $i$ is the vector $\bfw_i = (\bfw_{i1}, \bfw_{i2}, \dots, \bfw_{iJ_i})$. The complete observed data set is $\bfw = (\bfw_1, \dots, \bfw_n)$.  We write $\bfw_i\neq \mathbf{0}$ to indicate a capture history with length larger than zero, i.e. to represent the capture history of an individual observed at least once in the survey.

\subsection{Model formulation}
\label{s:likelihood_formulation}

Letting the model parameters be denoted by $\bftheta$, the likelihood is given by
\begin{equation*}
    L(\bftheta, N ; n,\bfw)\propto f(n ; \bftheta, N)f(\bfw ; n,\bftheta, N ),
\end{equation*}
where the first term corresponds to the marginal distribution of $n$ and the second term to the conditional distribution of the observed capture histories $\bfw$, given $n$ \citep{borchers2008spatially}. We let $p(\bftheta)$ denote the probability an individual is observed at least once in the survey, i.e. $p(\bftheta) = \mathbb{P}(\bfw_i\neq \mathbf{0})$. The first term is a binomial component,
\begin{equation*}
     f(n; \bftheta, N) = {N\choose n} p(\bftheta)^n (1-p(\bftheta))^{N-n}.
\end{equation*}
The second term, independent of $N$ given the number of observed individuals $n$, is given by
\begin{equation}
   f(\bfw ; n, \bftheta, N) = \prod_{i=1}^n f(\bfw_i ; \bfw_i\neq \mathbf{0}, \bftheta) = \prod_{i=1}^n \frac{f(\bfw_i; \bftheta)}{p(\bftheta)},
   \label{marg_likelihood}
\end{equation}
where the capture histories of distinct individuals are independent given the parameters. 

To formulate the terms $f(\bfw_i;\bftheta)$ and $p(\bftheta)$, we use the framework of a Markov-modulated marked Poisson process (MMMPP). We discretise the survey region as a grid, with each cell representing a state in the model. Individual movement is represented by a latent continuous-time Markov chain on this grid, where the probabilities of transitioning between states represent the movement dynamics of the animal. This state-space discretisation is a modelling choice that allows the model to be clearly expressed and fitted directly. Importantly, as the number of grid cells increases, the model approaches a continuous-space model. We describe how each component of the MMMPP is defined in the context of continuous-time SCR.

\vspace{-1 truemm}
 
\subsection*{State space $\mathcal{S}$}
The state space $\mathcal{S}=\{1,\dots,S\}$ is defined by discretising the landscape region $\mathcal{R}$ into $S$ square grid cells of equal area, such that each cell represents a geographic location, or state, in the model. Neighbouring cells are defined as those sharing an edge (up to four in this setting). We allow a cell to contain at most one camera.
The regular grid facilitates a clear and easily interpretable parameterisation of movement patterns between neighbouring cells. Since the computational cost scales with the number of states, a square-cell design provides a practical balance between spatial resolution and computational feasibility. The approach could be extended to include diagonal movements or to use irregularly shaped cells, which may better follow habitat boundaries (e.g., roads, rivers, or water bodies).

\vspace{-1 truemm}

\subsection*{Generator matrix $\bfQ$}
\label{G_matrix}

The generator matrix models the transition rates between states, which represents the movement behaviour of individuals across the landscape. The non-zero off-diagonal entries, $q_{rs}$, denote the rate at which an individual moves between cells $r$ and $s$. We permit movemen between only neighbouring cells (i.e., cells sharing an edge), so that for $r$ and $s$ that are not neighbours, $q_{rs}=0$. The diagonal elements are defined by $q_{rr} = -\sum_{s \ne r} q_{rs}$ to ensure that $\mathbf{Q}$ is a valid generator matrix with row sums equal to zero. We consider two formulations for $\bfQ$, representing different underlying ecological assumptions for animal movement:

\begin{description}
  \item[\textbf{(1) Random walk movement model (RWMove-SCR):}]

  This model assume spatially uniform movement rates, so that animals move via a random walk. We consider a continuous-time lattice approximation of Brownian motion  \citep{van1983stochastic}. For a regular square grid with mesh spacing $d$ (i.e. the distance between neighbouring cell centres), the transition rates for all neighboring pairs $(r,s)$ are defined as:
    $$
    q_{rs} = \frac{\sigma^2}{2d^2}, 
    $$
    where $\sigma^2$ is the diffusion coefficient parameter to be estimated. 
    Each (non-boundary) cell has four neighbours, so that the total rate of leaving a grid cell $r$ that is not located on the edge of the survey area is: 
    $$ q_{r} = \sum_{s\in neighbour(r)} \frac{\sigma^2}{2d^2} = \frac{2 \sigma^2}{d^2}. $$

    Larger $\sigma^2$ values correspond to more frequent transitions between cells and hence more dispersed movement; smaller values imply slower movement. As the number of grid cells increases, the model converges to a continuous-time Brownian motion with diffusion coefficient $\sigma^2$. In the continuous-time Markov formulation, the waiting time in state $r$ is exponentially distributed with rate $q_r$, with expected residence time in a cell is $1/q_r$.

\item[\textbf{(2) Ornstein-Uhlenbeck movement model (OUMove-SCR):}]

The second formulation is a discretised Ornstein–Uhlenbeck (OU) process \citep{uhlenbeck1930theory, miao2013analysis}, with a mean-reverting movement toward an individual-specific activity centre $\bfs_i$.
Transition rates are defined to match the first two moments of the OU process, such that for a state $r$ located at $(x_r, y_r)$, the corresponding rates to its four neighbouring cells are:

\begin{align*}
q_\text{right} &= \frac{\sigma^2}{2d^2} - \frac{\alpha(x_r - x_{s_i})}{2d}, &
q_\text{left}  &= \frac{\sigma^2}{2d^2} + \frac{\alpha(x_r - x_{s_i})}{2d}, \\
q_\text{up}    &= \frac{\sigma^2}{2d^2} - \frac{\alpha(y_r - y_{s_i})}{2d}, &
q_\text{down}  &= \frac{\sigma^2}{2d^2} + \frac{\alpha(y_r - y_{s_i})}{2d}.
\end{align*}
The parameter $\sigma^2$ controls the baseline movement intensity, while $\alpha \ge 0$ determines the strength of attraction towards the associated activity centre. The special case of $\alpha = 0$ reduces to the \emph{RWMove-SCR} model. The total leaving rate from cell $r$ is given by:
\begin{align*} 
q_r &= q_{\text{right}} + q_{\text{left}} + q_{\text{up}} + q_{\text{down}} = \frac{2\sigma^2}{d^2}.
\end{align*}
The intensity $q_r$ is independent of $\alpha$, so that while $\alpha$ influences directional bias it does not influence the overall movement rate. The expected residence time is again $1/q_r$.
\end{description}

\subsection*{Observation Matrix $\bfLambda$}

The observation matrix specifies the instantaneous Poisson detection rates for each state:
    $$\bfLambda=\begin{bmatrix} \lambda_1 & \cdots & 0 \\ \vdots & \ddots & \vdots \\ 
0 & \cdots & \lambda_S \end{bmatrix},$$
where $\lambda_s$ denotes the detection rate in cell $s \in \mathcal{S}$. We assume a common detection rate $\lambda_s = \lambda$ for states $s$ containing a camera trap; and $\lambda_s = 0$ otherwise (individuals can only be observed in cells with a camera trap). The state space is constructed such that each cell contains at most one camera. However, this framework can be extended to allow, for example, multiple cameras per cell; or for detection rates $\lambda_s$ to be a function of covariates to accommodate spatial heterogeneity in detectability (see Section \ref{MMMPP:discussion} for further discussion).

\subsection*{Mark distribution matrices $\mathbf{G}(\cdot)$}

We assume perfect spatial identification: a detection at trap $k$ implies that the individual occupies the same cell as the trap at that time. Consequently, the mark distribution matrix is diagonal with a single one in the element corresponding to the detected cell, and zeros elsewhere. This assumption represents the situation when grid cells match the spatial resolution of the camera traps, so that the detection range of a camera is fully contained within a single cell. This assumption can be relaxed to accommodate imperfect spatial information by specifying non-zero entries for multiple states, with detections providing probabilistic rather than deterministic information about latent location. These probabilities further permit distance-based detection functions while preserving the overall model structure.


\subsection*{Initial state distribution}

The initial state distribution $\boldsymbol{\pi}$ denotes the probability of an individual being in each state at time $t=1$. We consider two formulations: a uniform distribution; and the stationary distribution of the generator matrix.
The uniform distribution assumes an equal probability for all states and while  computationally simple to implement does not incorporate any spatial information on space utilisation. Alternatively, we may assume that individuals behave according to the stationary distribution of their movement dynamics. In this case $\boldsymbol{\pi}$ satisfies $\boldsymbol{\pi}\mathbf{Q}=\mathbf{0}$ and $\sum_{s=1}^S \pi_s=1$. For model  \emph{OUMove-SCR}, the stationary distribution depends on the individual-specific generator matrix. Since the state space is finite and connected, the generator matrix defines an irreducible continuous-time Markov chain, ensuring existence and uniqueness of the stationary distribution. 
We use the term \emph{stationary} here to emphasise the modeling assumption that populations have reached equilibrium at the start of the survey.
The choice between uniform and stationary initial distributions represents a trade-off between model simplicity and biological realism, with the stationary distribution generally more appropriate for established populations.

\subsection{Likelihood formulation}

We consider the likelihood contribution for each observed individual, $i=1,\dots,n$, using the HMM-based forward algorithm \citep{zucchini2009hidden}. The model parameters are $\bftheta = (\lambda, \sigma^2)$ for \emph{RWMove-SCR} and $\bftheta = (\lambda, \sigma^2, \alpha)$ for \emph{OUMove-SCR}. We focus on model \emph{OUMove-SCR}. We initially define the matrix exponential, 
\begin{equation*}
    \bfR(t,\bfs_i) = \exp(t(\bfQ(\bfs_i)-\bfLambda)).
\end{equation*}
The corresponding elements $r_{jk}(t,\bfs_i))$ denote the probability of an individual being in state $k$ at time $t$, given they are in state $j$ at time 0 and no detections occur, conditional on the activity centre $\bfs_i$. Thus, this involves integrating over both the continuous-time Markov chain dynamics in $\bfQ(\bfs_i)$ and the state-dependent Poisson observation process $\bfLambda$.

Define the time intervals $\tau_m = t_m - t_{m-1}$ for $m = 1, \dots, J_i$, with $t_0 = 0$. Conditioning on the activity centre of individual $i$, denoted $\mathbf{s}_i$, the corresponding likelihood contribution for individual $i$ is:
\begin{equation}
f(\bfw_i;\bftheta, \bfs_i) =  \bfpi (\bfs_i) \left( \prod_{m=1}^{J_i} \bfR(\tau_m, \bfs_i) \bfLambda \bfG (x_{ij}) \right) \bfR(T-t_{iJ_i}, \bfs_i) \boldsymbol{1}_S,
\label{e:Likelihood_stati}
\end{equation}
where $\boldsymbol{1}_S\in\mathbb{R}^S$ denotes the vector with all elements equal to one \citep{lu2012markov,mews2023markov}. The likelihood is composed of the product of the initial distribution of the state of the individual at the start of the survey; the probability of observing the set of capture times and locations (integrating over latent paths between observations, detection process and mark assignment); and the probability of not being observed following final capture.  

The probability of individual $i$ remaining unobserved for the duration of the survey period $T$ is given by: 
\begin{equation*}
 \bfpi(\bfs_i) \bfR(T,\bfs_i) \boldsymbol{1}_S.
\end{equation*}
Thus, the probability of being detected at least once in the survey is then the complement:
\begin{equation}
p(\bftheta ; \bfs_i) = 1 - \bfpi(\bfs_i) \bfR(T,\bfs_i) \boldsymbol{1}_S.
\label{e:proba_detected}
\end{equation}
Equations (\ref{e:Likelihood_stati}) and (\ref{e:proba_detected}) are conditional on the individual-specific latent activity centre and so for the observed likelihood we need to integrate these out. Assuming that the activity centres are uniformly distributed over the region $\mathcal{R}$ of area $A$, we obtain: 
\begin{equation}
\label{e:ind_lik_integral}
    f(\bfw_i; \bftheta) = \int_{\mathcal{R}} f(\bfw_i \mid \bfs, \bftheta) \frac{1}{A} d\bfs,
\end{equation}
\begin{equation}
\label{e:obs_int}
    p(\bftheta) = \int_{\mathcal{R}}  p(\bftheta \mid \bfs)  \frac{1}{A} d\bfs.
\end{equation}
Substituting Equations (\ref{e:ind_lik_integral}) and (\ref{e:obs_int}) into Equation (\ref{marg_likelihood}) yields the observed data likelihood. 

The corresponding likelihood for model \emph{RWMove-SCR} follows similarly as well as alternative choices of initial distribution. We note that for \emph{RWMove-SCR} model, where all individuals share the same generator matrix $\bfQ$, no spatial integration is needed.

\section{Model fitting} 
\label{s:model fitting}

We adopt a conditional likelihood approach where we obtain the MLEs of the parameters from Equation~(\ref{marg_likelihood}). Recall, for model \emph{RWMove–SCR}, $\bftheta = (\sigma^2,\lambda)$; while for model \emph{OUMove–SCR}, $\bftheta = (\sigma^2,\alpha,\lambda)$. The total population size $\widehat{N}$ is estimated via a Horvitz–Thompson‐type estimator \citep{horvitz1952generalization},
\begin{equation*}
    \widehat{N} = \frac{n}{p(\widehat{\bftheta})}.
\end{equation*}
The associated variance can be computed following \cite{huggins1989statistical} and  \cite{alho1990logistic}.

\subsection*{Computational considerations}

To fit the \emph{Move-SCR} model, we used the Template Model Builder (\texttt{TMB}) package \citep{kristensen2016tmb} to write code in \texttt{C++} and optimise the parameters in \texttt{R} \citep{R}. \texttt{TMB} provides automatic differentiation for efficient gradient-based optimisation, making it well-suited for computationally intensive likelihood evaluations such as those required here.

Evaluating the likelihood requires repeated computations of the $S \times S$ matrix exponential:
\
\[
\bfR(\tau) = \exp\big(\tau(\bfQ - \bf\Lambda)\big).
\]
The matrix exponential is defined through an infinite power series, 
\[
\exp(\bfM) = \sum_{r=0}^\infty \frac{1}{r!} \bfM^r,
\]
which implicitly sums over all possible movement paths between states. In general, there is no closed form expression for $\exp(\bfM)$ and numerical approximations are required \citep{moler1978nineteen,moler2003nineteen}. To evaluate the matrix exponentials we use the \textit{atomic::expm} function in \texttt{TMB}, which implements a scaling–and–squaring algorithm combined with a high-order Padé approximation \citep{higham2005scaling,higham2008functions}. This method
is numerically robust but computationally intensive, with a complexity of order $\mathcal{O}(S^3)$ per matrix exponential.

The integrals over unknown activity centre in Equations (\ref{e:ind_lik_integral}) and (\ref{e:obs_int}), are performed following standard SCR practices \citep{borchers2008spatially}. The \emph{OUMove–SCR} model is more computationally demanding, with the matrix $\mathbf{Q}$ depending on the individual-specific activity centres and must be recomputed for each evaluation. Nevertheless, implementation remains feasible for moderately large state spaces.

\section{Simulation Study}
\label{MMMPP:simulations}

We conduct a simulation study to evaluate the performance of the \emph{Move-SCR} 
framework.
We define the state space to be a $10 \times 10$ grid of 100 square cells, each with a side length of 1~km, giving a total study area of 100~km$^2$. 
We set the survey duration to $T = 11$ days and the true population size to $N = 20$, chosen to reflect conditions similar to the case study presented in 
Section~\ref{MMMPP:application}. Camera traps are placed in 30 randomly selected cells at the start of each replicate, excluding a buffer of 2~km around the boundary, to avoid edge effects. We simulate 100 replicate data sets.

\subsection*{Simulation procedure}

Data are generated from the \emph{OUMove-SCR} model as follows. For each replicate:

\begin{enumerate}
\item Sample 30 cells (without replacement) uniformly from the interior of the grid as camera trap 
    locations.
\item For each individual $i=1,\dots,N$:
\begin{enumerate}
\item Sample an activity centre $\bfs_i$ uniformly over the survey region and 
        construct the individual-specific generator $\bfQ(\bfs_i)$. 

        \item Sample an initial state $x_0$ from the stationary distribution 
        $\bfpi(\bfs_i)$, obtained by solving $\bfpi(\bfs_i)\bfQ(\bfs_i) = \bfzero$ subject to $\sum_j \pi_j(\bfs_i) = 1$.

        \item Simulate movement as a continuous-time Markov chain: residence time in state $j$ is drawn 
        from $\mathrm{Exp}(-Q_{jj}(\bfs_i))$, and transitions are sampled according to the normalised off-diagonal entries of row $j$ of $\bfQ(\bfs_i)$.

        \item Generate detections via a Poisson process such that in each interval spent in state $j$, detections occur at rate $\lambda_j = \lambda \cdot 
        \mathbf{1}_{[\text{state } j \text{ contains a trap}]}$, with detection times uniform over the interval. 
\end{enumerate}
\item Retain only individuals with at least one detection; let $n$ denote the 
    number of observed individuals.
\end{enumerate}

We set $\lambda = 0.5$ and $\sigma^2 = 1$, and simulate 100 replicate data sets for each of $\alpha \in \{0, 0.5, 1\}$. On average, 12 out of 20 individuals were observed per data set. The mean number of detection events increased with $\alpha$, from 33.4 at $\alpha = 0$ to 41.8 at $\alpha = 1$. Each data set is fitted with \emph{RWMove-SCR}, \emph{OUMove-SCR}, and the continuous-time SCR model which we refer to as CT-SCR. CT-SCR is formulated and fitted following \cite{borchers2014continuous}. For the \emph{OUMove-SCR} and CT-SCR models, the activity centres are integrated out over the same grid as the state space. 

\subsection*{Simulated trajectories}

\begin{figure}[!htbp]
    \centering
        \includegraphics[width=\textwidth]{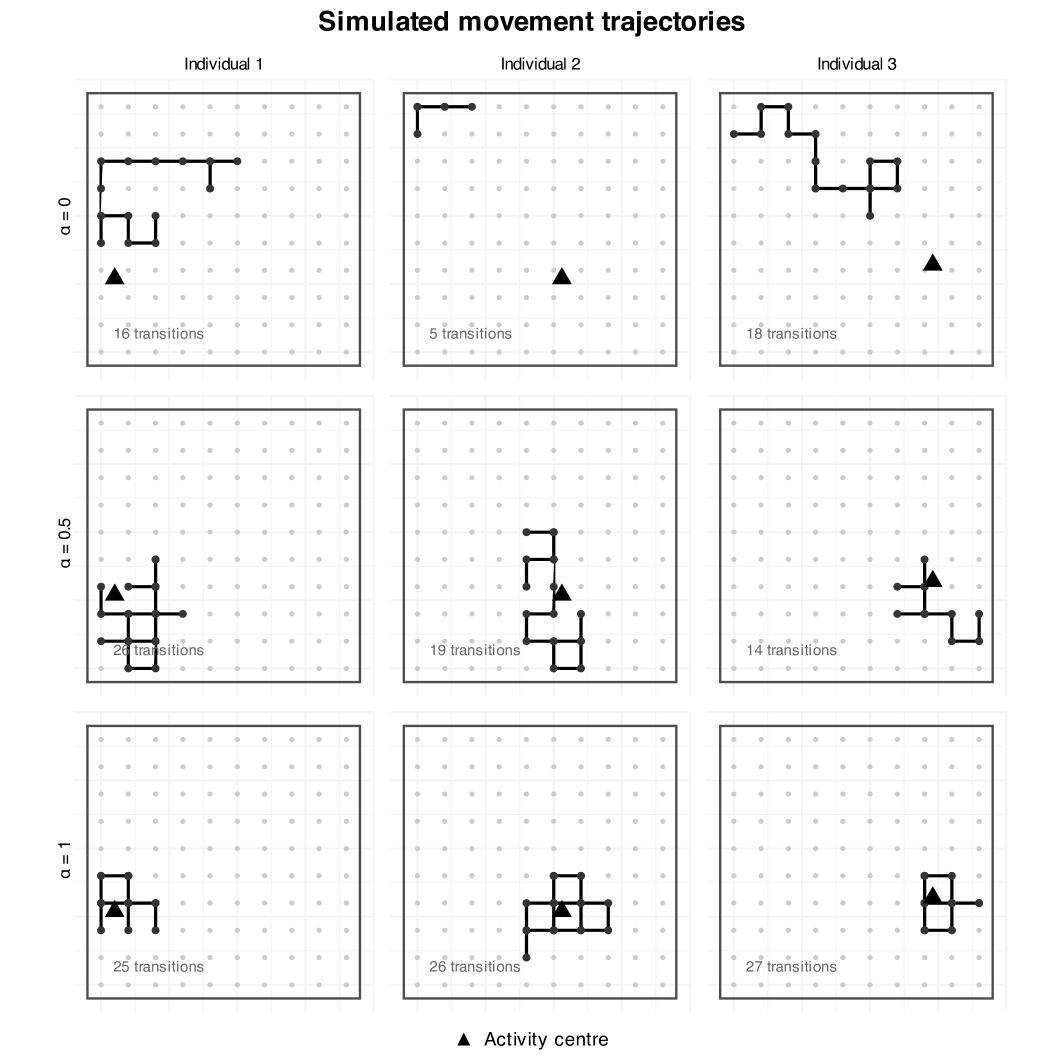}
    \caption{Simulated movement trajectories for three individuals under the OUMove-SCR model with $\sigma^2 = 1$ over $T = 11$ days, for three values of the attraction parameter $\alpha \in \{0, 0.5, 1\}$ (rows). The state space is a square grid of $10\times10$ cells, with each grey dot representing the centre of a cell. The activity centres are located at the black triangles. Each column of graphs corresponds to one individual, whose activity centre ($\blacktriangle$) was drawn randomly from the landscape. The number of transitions between grid cells is shown in each panel.}
    \label{fig:trajectories}
\end{figure}

Figure~\ref{fig:trajectories} shows nine simulated movement trajectories for model \emph{OUMove-SCR}, for three individuals for $\alpha \in \{0, 0.5, 1\}$. When $\alpha = 0$, the model reduces to a random walk, so trajectories explore the state space without any tendency to return to the activity centre. As $\alpha$ increases, individuals spend progressively more time near their activity centre: at $\alpha = 0.5$ animals explore a moderate range, while at $\alpha = 1$ trajectories are more tightly concentrated. The number of transitions, shown in each panel, illustrates that an individual visiting less grid cells does not necessarily change state less often.

\subsection*{Results}

Table~\ref{t:simulations} summarises model performance for different values of $\alpha$, and Figure~\ref{fig:boxplot} shows the distribution of $\widehat{N}$ estimates. \emph{OUMove-SCR} achieves the lowest bias and RMSE for $\widehat{N}$ across all scenarios, as expected given that data are generated from this model. \emph{RWMove-SCR} performs comparably for $\alpha = 0$, where the two models are theoretically equivalent but differ in implementation (see Appendix~A), yet shows increasing bias as $\alpha$ increases, reflecting misspecification of the movement process. CT-SCR consistently overestimates $N$ and exhibits the largest RMSE across all scenarios, with bias increasing as $\alpha$ increases, indicating that not accounting for the movement structure leads to systematic overestimation of population size.

\emph{RWMove-SCR} and CT-SCR were computationally fast across all scenarios, with mean times of under 10 and 25 seconds per data set, respectively. \emph{OUMove-SCR} was more computationally expensive, with mean times of approximately 20 minutes at $\alpha = 0$, increasing to 33 minutes for $\alpha = 0.5$ and $\alpha = 1$. The computational cost of \emph{OUMove-SCR} relative to \emph{RWMove-SCR} and CT-SCR results from the multiple matrix exponential calculations required to integrate out the activity centre.
Additional details on data set characteristics, computation times, and the differences between \emph{OUMove-SCR} and \emph{RWMove-SCR} in simulations with $\alpha = 0$ are provided in Supplementary Appendix A. A sensitivity analysis relating to the state-space discretisation are provided in Supplementary Appendix B.

\begin{table}[!htbp]
\centering
\small
\setlength{\tabcolsep}{6pt}
\renewcommand{\arraystretch}{1.15}
\begin{tabular}{l l r r r}
\toprule
\textbf{Model} & \textbf{Metric} & \(\alpha=0\) & \(\alpha=0.5\) & \(\alpha=1\) \\
\midrule
\multirow{4}{*}{\emph{OUMove-SCR}}
 & Mean \(\widehat N\) (SE) & 19.3 (3.89) & 21.3 (4.09) & 21.7 (4.29) \\
 & \% Bias             & -3.66 & 6.35 & 8.55 \\
 & RMSE                & 4.29  & 5.28   & 5.55   \\
\midrule
\multirow{4}{*}{\emph{RWMove-SCR}}
 & Mean \(\widehat N\) (SE) & 21.0 (4.27) & 23.9 (4.6) & 23.4 (6.50) \\
 & \% Bias             & 4.99 & 16.45  & 17.2   \\
 & RMSE                & 4.58  & 5.64   & 6.02   \\
\midrule
\multirow{4}{*}{CT-SCR}
 & Mean \(\widehat N\) (SE) & 22.0 (5.20) & 24.2 (5.27) & 25.6 (5.76) \\
 & \% Bias             & 10.1 & 21.0  & 28.2  \\
 & RMSE                & 5.84 & 6.52  & 7.72  \\
\bottomrule
\end{tabular}
\caption{ Simulation study results. Data were generated from the \emph{OUMove-SCR} model with $T = 11$ days, $N = 20$, $\lambda = 0.5$, and $\sigma^2 = 1$, across 100 replicate data sets for each value of $\alpha \in \{0, 0.5, 1\}$. Shown are the mean estimated $\widehat{N}$ with its standard error (SE), percentage bias, and root mean squared error (RMSE).}
\label{t:simulations}
\end{table}

\begin{figure}[!htbp]
    \centering
        \includegraphics[width=\textwidth]{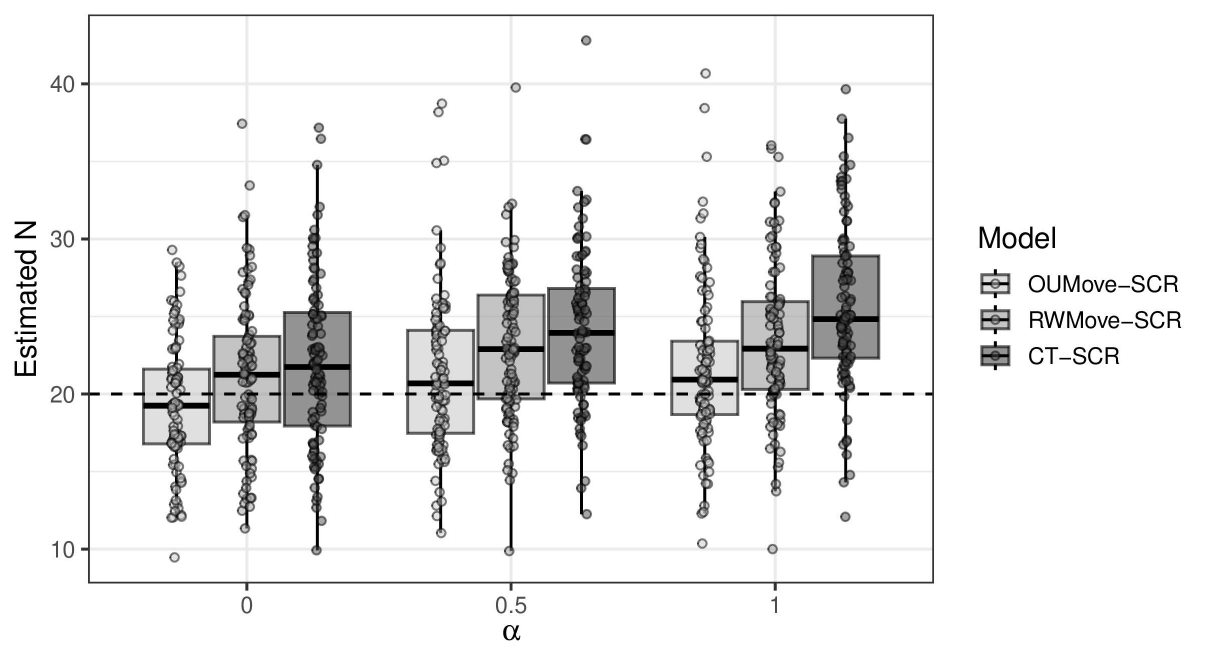}
    \caption{Estimated population size ($\widehat{N}$) across 100 replicate data sets simulated from the \emph{OUMove SCR} model for $\alpha\in {0,0.5,1}$. Dashed lines indicate the true population size $N=20$.}
    \label{fig:boxplot}
\end{figure}

\section{Application: American Martens}
\label{MMMPP:application}

\begin{figure}[!htbp]
    \centering
\includegraphics[width=0.7\textwidth]{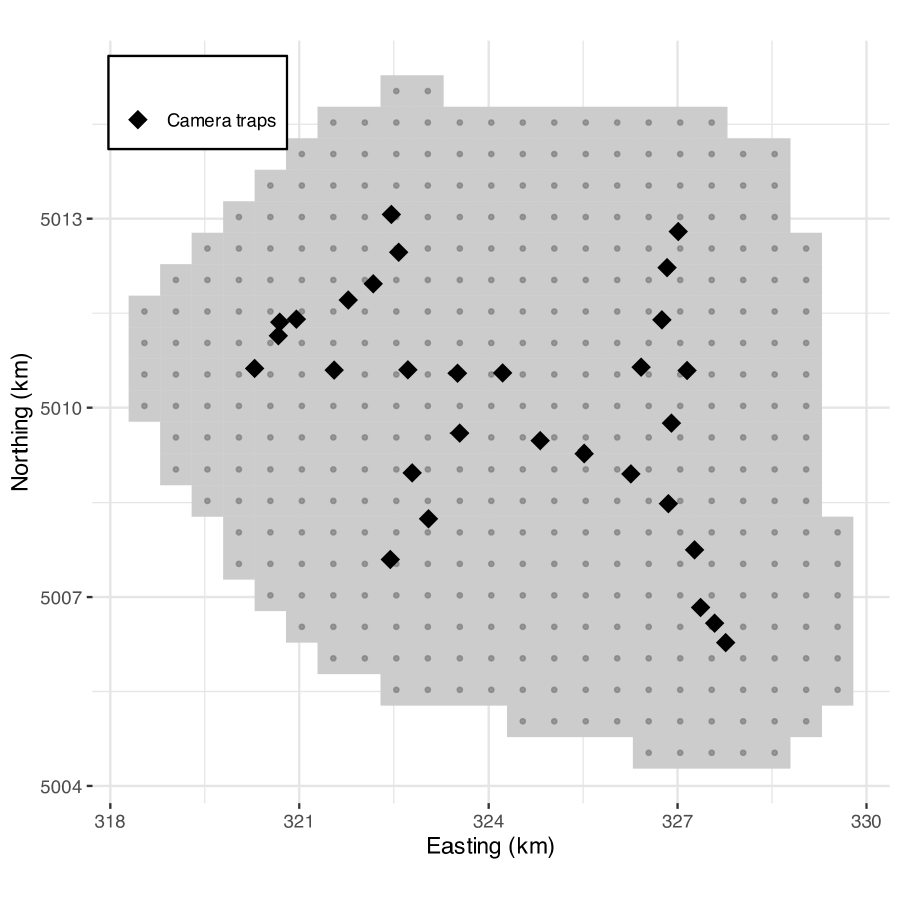}
    \caption{Study area of the American marten survey (units in kilometers). The state space consists of $S=381$ grid cells, with grey dots indicating the centre of each cell. Camera trap locations are shown as black diamonds. The state space was constructed by adding a 2-km buffer to the convex hull formed by the camera traps.
}
 \label{f:landscape}
\end{figure}

We consider camera trap data of American martens in New Hampshire, USA. The survey was conducted from 28 March to 7 April 2017 (11 days), and 30 camera traps were active throughout this time frame. Individual martens were uniquely recognised from their chest markings. A total of 9 individuals were observed, with 8.22 detections on average per individual (median 6, maximum 28), with 74 detections in total. 
To define the study area, we added a buffer of 2-km to the area enclosed by the convex hull of the camera traps, following standard SCR methods \citep{efford2004density,borchers2008spatially,royle2013spatial} and knowledge of American martens \citep{dumyahn2007winter}. The resulting state space and trap layout is displayed in Figure \ref{f:landscape}, with a total area $A$ of 95.25~km\(^2\). 
The multiple observations per individual over the short survey period would suggest there to be temporal and spatial dependence. 
We fit both the \emph{RWMove-SCR} and the \emph{OUMove-SCR} models and estimate the corresponding parameter vector $\widehat{\bftheta}$. 

We define a state space of size $S=381$, corresponding to a regular grid with cell width 0.5~km (i.e., 0.25~km\(^2\) per cell) with no grid cell containing more than one camera. We assume that an individual detected by a camera is located in the same grid cell as the camera at the time of observation. 
The integration mesh over the activity centre locations used to fit \emph{OUMove-SCR} contained 41 grid cells, a pragmatic balance between computational feasibility and model accuracy.

Table \ref{t:Americanmartens_MMMPP} provides the results of models \emph{RWMove-SCR}, \emph{OUMove-SCR} and CT-SCR. The models took $8.06$ minutes to fit for \emph{RWMove-SCR}, 3.37 hours for \emph{OUMove-SCR}, and 2.95 minutes for CT-SCR. 
Both \emph{Move-SCR} models showed substantially lower AIC values than CT-SCR, indicating improved model fit, while little difference in AIC was observed between \emph{RWMove-SCR} and \emph{OUMove-SCR}. The estimated population size is 14.40 (3.09) for CT-SCR, increasing slightly to 15.44 (3.66) for  \emph{OUMove-SCR} and to 16.03 (3.60) for \emph{RWMove-SCR}. The confidence intervals indicate a substantial overlap of the population estimates for all three models.

\begin{table}[htbp]
  \centering
  \begin{subtable}[t]{\textwidth}
    \centering
    \begin{tabular}{ccc}
      \toprule
      \multicolumn{3}{c}{\textbf{\emph{RWMove-SCR} model}} \\
      \multicolumn{3}{c}{AIC =  293.22} \\
      \midrule
      Parameter & Estimate (SE) & 95\% CI \\ 
      \midrule
      $N$ & 16.03 (3.60) & (8.97, 23.09) \\ 
      $\sigma^2$  & 0.38 (0.080) & (0.22, 0.54)  \\
      $\lambda$  & 3.81 (0.61) & (2.61, 5.00) \\
      \bottomrule
    \end{tabular}
    \caption{Parameter estimates from fitting the \emph{RWMove-SCR} model to the American marten data set.}
    \label{t:American_MMMPPa}
  \end{subtable}  
  
  \vspace{1em}

  \begin{subtable}[t]{\textwidth}
    \centering
    \begin{tabular}{ccc}
      \toprule
      \multicolumn{3}{c}{\textbf{\emph{OUMove-SCR} model}} \\
      \multicolumn{3}{c}{AIC = 293.73 } \\
      \midrule
      Parameter & Estimate (SE) & 95\% CI \\ 
      \midrule
      $N$ & 15.44 (3.66) & (8.27, 22.61) \\ 
     $\sigma^2$  & 0.40 (0.090) & (0.23, 0.58)  \\
      $\alpha$  & 0.37 (0.47) & (-0.56, 1.29) \\
      $\lambda$ & 3.86 (0.62) & (2.64, 5.07) \\
      \bottomrule
    \end{tabular}

    \caption{Parameter estimates from fitting the OUMove-SCR model to the American marten data set.}
    \label{t:American_MMMPPb}
  \end{subtable}

   \vspace{1em}
  
  \begin{subtable}[t]{\textwidth}
    \centering
    \begin{tabular}{ccc}
      \toprule
      \multicolumn{3}{c}{\textbf{CT-SCR Model}} \\
      \multicolumn{3}{c}{AIC = 421.78} \\
      \midrule
      Parameter & Estimate (SE) & 95\% CI \\ 
      \midrule
      $N$ & 14.40 (3.09) & (8.34, 20.46) \\ 
      $h_0$ & 1.47 (0.55) & (0.71, 3.04) \\
      $\sigma_{SCR}^2$ & 0.27 (0.050) & (0.18, 0.39) \\
      \bottomrule
    \end{tabular}

    \caption{Parameter estimates from fitting the CT-SCR model to the American marten data set.}
    \label{t:American2b}
  \end{subtable}

  \caption{Parameter estimates, standard errors (SE) and 95\% confidence intervals (CI) from fitting the \emph{RWMove-SCR} and \emph{OUMove-SCR} models to the data obtained from the American martens survey. $S$ is set to 381, and the activity centre integration was performed on a mesh of 41 points.}
  \label{t:Americanmartens_MMMPP}  
\end{table}

Using the expression for the transition rates in the \emph{RWMove-SCR} model (Section~\ref{G_matrix}) with grid spacing $d=0.5$ km and estimated diffusion parameter $\widehat{\sigma}^2 = 0.38$ (SE $= 0.080$), the transition rate between neighbouring cells $r$ and $s$ is estimated as $\widehat{q}_{rs} = 0.76$. Assuming that $r$ is not on the boundary of the survey area, the total leaving rate is $\widehat{q}_r=3.04$, yielding an expected residence time of 0.33 days ($\approx$ 7.89 hours). Propagating uncertainty from $\sigma^2$ using the delta method gives an approximate standard error of 1.73 hours and a $95\%$ confidence interval of $(4.49, 11.30)$ hours. This range remains consistent with field-based estimates of marten movement \citep{moriarty2016forest}.

For the \emph{OUMove-SCR} model, using the estimated parameters $\widehat{\sigma^2} = 0.40$ (SE = 0.088) and $\widehat{\alpha} = 0.38$ (SE = 0.47), the total leaving rate is $\widehat{q_r} = 3.20$. The corresponding expected residence time in a cell is $\frac{1}{3.20} =  0.3125 \text{ days (}\approx\text{7.50 hours)}$, with a standard error of 1.66 hours and a 95$\%$ confidence interval of (4.27,10.73) hours. Despite the inclusion of directional bias in the \emph{OUMove-SCR} model, the estimated residence times have substantially overlapping confidence intervals with those of the \emph{RWMove-SCR} model, indicating comparable overall movement rates across the two formulations.

The estimate of the detection rate, $\lambda$, is very similar for both models: $\widehat{\lambda} = 3.81$ (SE = 0.61) for \emph{RWMove-SCR} and $\widehat{\lambda} = 3.86$ (SE = 0.62) for \emph{OUMove-SCR}, corresponding to an expected number of detections of $41.91$ and $42.46$ over the $T = 11$ day survey period, respectively, for an individual remaining in a cell containing a camera trap throughout.

The attraction strength parameter of the \emph{OUMove-SCR} model is estimated to be $\widehat{\alpha} = 0.38$ (SE = 0.47). The large standard error with a confidence interval including zero suggests that the movement of American martens in this data set is not greatly informed by attraction to activity centres, a conclusion supported by $\Delta$AIC = 0.51 in favour of \emph{RWMove-SCR}.

\section{Discussion}
\label{MMMPP:discussion}

We propose a novel \emph{Move–SCR} framework, providing a continuous-time extension of spatial capture–recapture models that explicitly incorporates animal movement via MMMPPs. By formulating movement as a continuous-time Markov chain over a discretised landscape and modeling detections as state-dependent Poisson events, the framework accommodates irregular detection times while jointly estimating population abundance and movement dynamics. This approach builds on continuous-time SCR formulations \citep{borchers2014continuous} and provides a continuous-time analogue to discrete-time hidden Markov models incorporating movement into SCR \citep{crum2023forecasting}. In contrast to recent movement-integrated SCR approaches \citep[e.g.,][]{converse2022special, mcclintock2022integrated, gardner2022integrated}, \emph{Move–SCR} offers a likelihood-based approach that does not require auxiliary telemetry data.

The MMMPP-based formulation unifies movement and detection processes, enabling inference on both the latent movement patterns and population parameters. Application to American marten data illustrates the practical implementation of two variants: the uniform movement model (\emph{RWMove–SCR}) and the activity centre–guided model (\emph{OUMove–SCR}). The framework is also flexible with respect to the observation process. For example, multiple cameras could be accommodated within a single grid cell, and state-specific detection rates $\lambda_s$ could be modeled as functions of spatial or environmental covariates to account for heterogeneity in detectability. Extensions to time-varying covariates are however not straightforward and would require alternative numerical approaches, as the current matrix exponential formulation would no longer be applicable.

Continuous-time modeling offers several advantages over discrete-time approaches. Firstly, it preserves the natural temporal resolution of detection data without aggregation into arbitrary survey occasions. Secondly, it allows flexible representation of movement at arbitrary spatial scales through discretisation of the landscape, with the discrete representation converging to continuous-space movement as the grid resolution increases. Simulation studies suggest that the framework is robust to moderate discretisation choices, though overly-coarse grids can induce bias and reduced coverage. Simulation results highlight a trade-off between model complexity and computational cost: \emph{RWMove–SCR} provides stable inference with fewer parameters, whereas \emph{OUMove–SCR} improves performance when home-ranging behaviour is present and supported by the data.

Evaluating matrix exponentials and integrating over latent activity centres remains the primary computational challenge  of the model. Nevertheless, exploiting sparsity in the generator matrix, concentrating integration points near observed detections, and parallel computation are practical strategies that mitigate these challenges. Future research could explore alternative numerical algorithms and computational techniques to further reduce these costs.

Further areas of current interest include extending the approach to irregular grid definitions of the state space, allowing individuals to be detected in cells outwith those containing camera traps, alternative movement models such as correlated random walks or velocity–jump processes \citep{jonsen2005robust, johnson2008continuous, codling2005calculating}, and the incorporation of time-varying or spatial covariates in both movement and detection processes.

In conclusion, the \emph{Move–SCR} framework provides a flexible likelihood-based approach for joint modeling of movement and spatial capture–recapture in continuous time. By embedding movement directly within the likelihood while preserving temporal resolution, \emph{Move–SCR} broadens the class of SCR models that can be fit without auxiliary telemetry data. Improving computational efficiency and extending the class of movement models remain important directions for future research.



\section*{Acknowledgements}

We would like to thank Donovan Drummey, Jill Kilborn, and Chris Sutherland for providing access to the American marten spatial capture-recapture data. The collection of this data was supported by funding from the New Hampshire Fish and Game Department and the University of Massachusetts-Amherst. We greatly thank Paul Blackwell for the valuable discussions that contributed to the development of this work.


For the purpose of open access, the author has applied a Creative Commons Attribution (CC BY) licence to any Author Accepted Manuscript version arising from this submission.

\section*{Supplementary Material}

The Supplementary Material contains additional simulation results and supporting figures. Appendix A reports further results from the simulation study, and Appendix B presents an additional investigation of the effects of spatial discretisation. The code used in this manuscript for the simulations and the analysis of the American marten data is available on GitHub: \url{https://github.com/clarapasu/Move-SCR}.

\section*{Data Availability Statement}

The American marten data supporting the findings in this paper are available on GitHub at \newline \url{https://github.com/clarapasu/Move-SCR}.

\vspace*{-8pt}

\bibliographystyle{biom} 
\bibliography{refs}

\label{lastpage}

\clearpage
\appendix
\section*{Appendix A: Additional results from the Simulation study}

We present additional results from the simulation study described in the main text. Table~\ref{t:data_summary} summarises the average number of observed individuals, the number of detection events, and the computational time across the simulated data sets for each value of $\alpha$.

\begin{table}[!htbp]
\centering
\begin{tabular}{r r r r r r}
\toprule
$\alpha$ & Mean $n_{\text{obs}}$ & Mean $n_{\text{events}}$ & 
\multicolumn{3}{c}{Mean run time (s)} \\
\cmidrule(lr){4-6}
 & & & \emph{RWMove-SCR} & \emph{OUMove-SCR} & CT-SCR \\
\midrule
0.0 & 11.9 & 33.4 & 6.3  & 1171 & 21.0 \\
0.5 & 12.5 & 41.0 & 7.5  & 1990 & 24.3 \\
1.0 & 12.2 & 41.8 & 7.7  & 2002 & 24.8 \\
\bottomrule
\end{tabular}
\caption{Mean number of observed individuals, detection events, and computation 
time per dataset across 100 simulations, by value of $\alpha$.}
\label{t:data_summary}
\end{table}

Figure~\ref{fig:boxplot} presents parameter recovery results for the scenario where $\beta = 0$, including estimates of the detection parameter $\lambda$ and the movement parameters $\sigma^2$ and $\alpha$. Estimates of $\lambda$ are well recovered under both \emph{OUMove-SCR} and \emph{RWMove-SCR}, with negligible bias. 

Under \emph{OUMove-SCR}, the parameter $\alpha$ is constrained to be positive, such that the true value $\alpha = 0$ lies on the boundary of the parameter space. Consequently, the model struggles to recover values close to zero exactly, and estimates are instead slightly shifted upward, here around $0.1$. Estimates of $\sigma^2$ under \emph{OUMove-SCR} also show a slight positive bias relative to the true value of $1$, which may partially compensate for the overestimation of $\alpha$ in reproducing the overall movement dynamics.

\begin{figure}[!htbp]
    \centering
        \includegraphics[width=\textwidth]{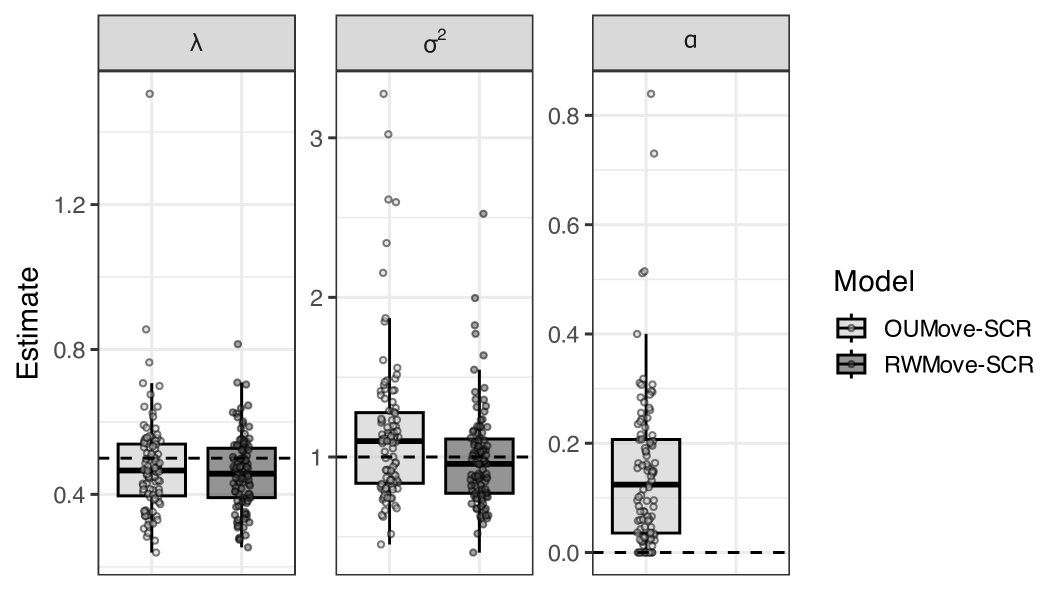}
    \caption{test}
    \label{fig:boxplot}
\end{figure}

These results show that even when $\beta=0$, \emph{OUMove-SCR} does not exactly simplify to \emph{RWMove-SCR}, as the likelihoods differ numerically: \emph{OUMove-SCR} integrates over activity centres for unobserved individuals on a mesh, whereas \emph{RWMove-SCR} uses a uniform movement kernel without that integration. These implementation differences, combined with the variability given by the $100$ replicates, yield small but systematic differences between the models. In the absence of an activity-centre effect, the simpler \emph{RWMove-SCR} model recovers $\lambda$ and $\alpha$ slightly more accurately than \emph{OUMove-SCR}, which estimates an additional, unnecessary parameter.

\section*{Appendix B: Effect of Discretisation Mismatch}

The \emph{Move-SCR} framework approximates continuous animal movement using a continuous-time Markov chain defined over a discrete spatial grid. Although animal movement occurs in continuous space and time, movement is represented through transitions between neighbouring spatial states. Theoretically, as grid-cell size approaches zero, the model converges to a continuous-space movement process. In practice, however, computational constraints limit the number of states $S$ that can be considered, meaning that the chosen discretisation may not fully match the true spatial scale of movement. Understanding the impact of this discretisation choice is therefore important both for model accuracy and computational feasibility.

\begin{figure}[!htbp]
    \centering
    \includegraphics[width=0.7\textwidth]{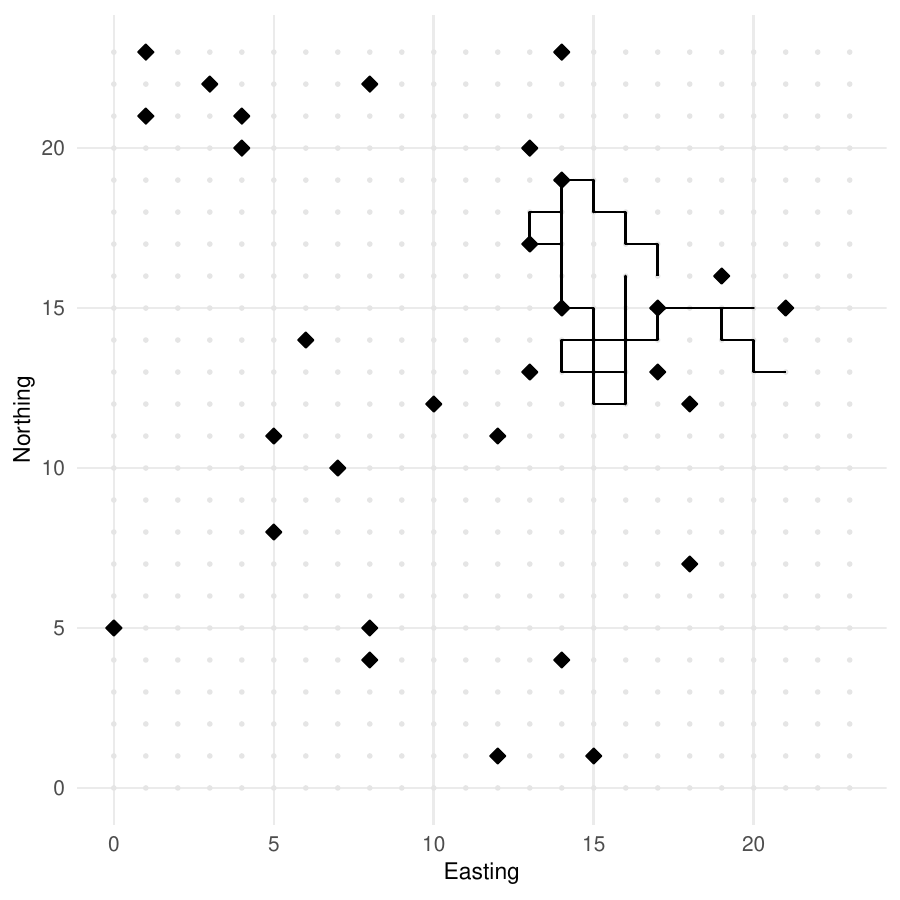}
    \caption{Example simulated trajectory under the \emph{RWMove-SCR} model. Grey points represent the centres of spatial states and black diamonds indicate camera-trap locations. An individual may be detected when occupying a state containing a trap. Distances are expressed in kilometres.}
    \label{f:trajectory_example}
\end{figure}

We investigate the robustness of the \emph{RWMove-SCR} model to different spatial discretisations. Data are generated from the \emph{RWMove-SCR} model using a fine spatial discretisation, $S=576$, chosen to approximate continuous movement and produce realistic detection patterns. The simulation parameters are set as $\alpha = \log(0.05)$, $\lambda = 0.10$, $N = 30$ individuals and $T = 11$ days. These values yield an average number of $15$ observed individuals and $41.8$ detections per data set. An example trajectory is displayed in Figure~\ref{f:trajectory_example}.

 We generate 50 data sets from this set up and fit the model under five different discretisation levels: $S \in \{49, 81, 144, 289, 400\}$. This design allows the evaluation of both the impact of discretisation mismatch on parameter recovery and the associated computational efficiency. The aim of this analysis is not to identify a universally optimal discretisation, but rather to illustrate the trade-off between computational cost and approximation accuracy within the \emph{Move-SCR} framework.

\begin{table}[htbp]
  \centering
  \small
  \renewcommand{\arraystretch}{1.15}
  \begin{tabular}{lccccc}
    \toprule
    State space size & \textbf{$S=49$} & \textbf{$S=81$} & \textbf{$S=144$} & \textbf{$S=289$}   & \textbf{$S=400$} \\
    \midrule
    $\widehat{N}$ (SE)    & 21.84  (3.10) & 22.37 (3.26) & 24.35 (3.85) & 27.72 (4.80) & 29.71 (5.35)  \\
    Bias $N$       &  -8.16 & -7.63 & -5.65 & -2.28 & -0.29 \\
    Coverage $N$ (\%)    &  38 & 42 & 58 & 78 & 84   \\
    Runtime (sec)               & 1.04 & 4.07 & 21.9 & 141 & 669       \\
    RMSE N   &  9.50 & 9.16 & 7.70 & 6.85 & 7.16     \\
    \bottomrule
  \end{tabular}
\caption{Results of the performance of the RWMove-SCR model under different spatial resolutions. Data were simulated using a state space size of $S = 576$. The results of this table are the mean across 50 replicates with $N = 30$, $T=11$ days, $\alpha=log(0.05)$ and $\lambda=0.1$. The results and metrics include the parameter estimate $\widehat{N}$, standard error, bias, coverage as proportion of $ 95\%$ CIs containing true value (Coverage $N$ (\%)), average runtime of the model in seconds, and RMSE.}
  \label{t:misspecification}
\end{table}

Table~\ref{t:misspecification} summarises the model performance across the different state space discretisations. Population size is underestimated under all discretisations, although estimates increase toward the true value as the number of states increases and the discretisation becomes finer. Mean estimates range from $21.84$ for $S = 49$ to $29.71$ for $S = 400$. Similarly, the bias decreases steadily in magnitude, from $-8.16$ for the coarsest discretisation to $-0.29$ for the finest.

The RMSE is largest for small state spaces and stabilises around 7 for larger values of $S$. Coverage of the nominal 95\% confidence intervals also improves substantially with increasing state space resolution, increasing from 38\% for $S = 49$ to 84\% for $S = 400$. Although the standard error of $\widehat{N}$ increases with $S$, reflecting the greater complexity of the fitted model, the resulting estimates are substantially less biased.

\begin{figure}[ht]
        \centering
        \includegraphics[width=0.7\textwidth]{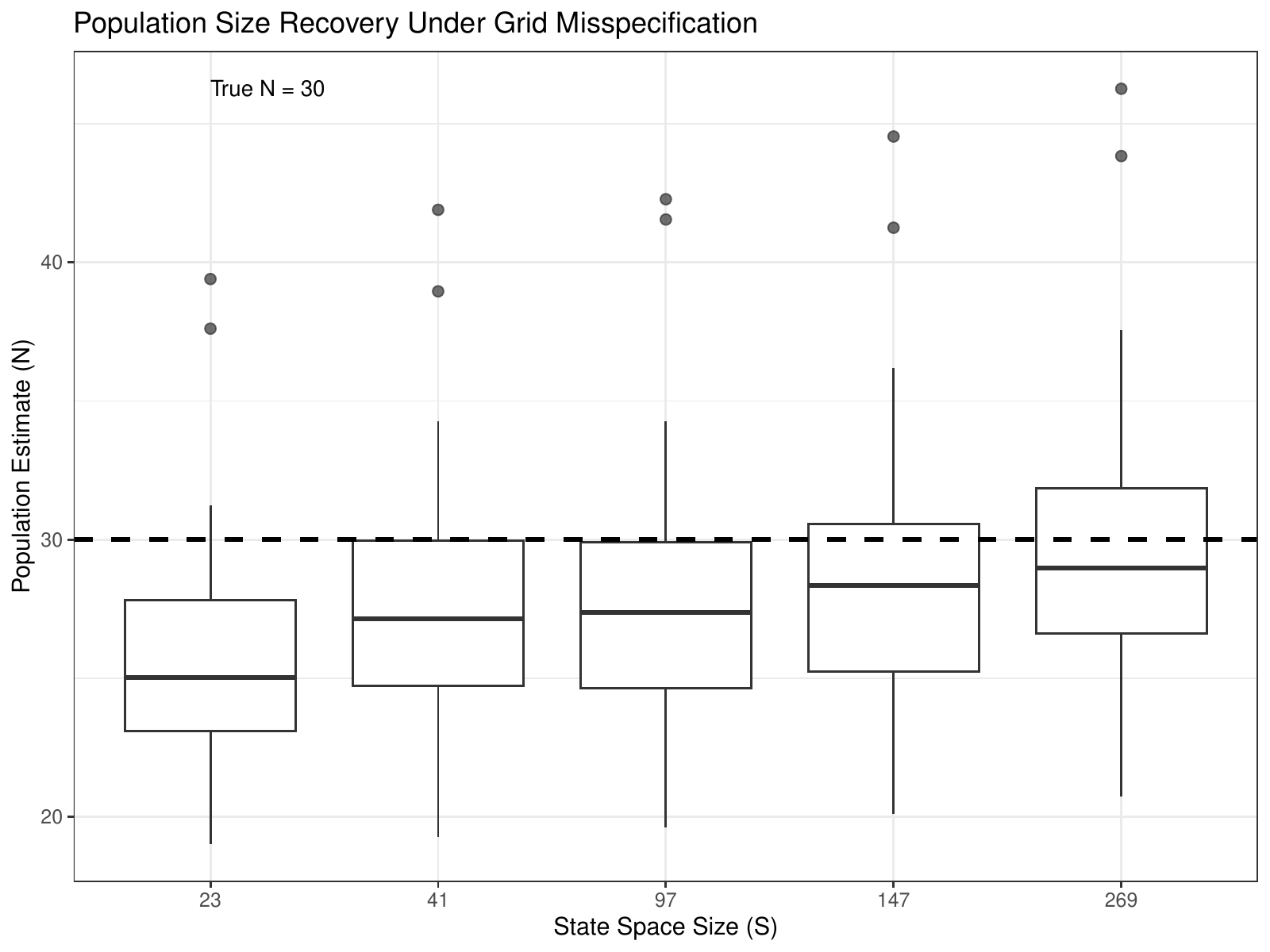}
\caption{Population size estimates from the \emph{RWMove-SCR} model across 50 simulated data sets generated with state space size $S = 576$ and fitted using different spatial discretisations. The dashed horizontal line indicates the true population size, $N = 30$.}
    \label{f:sim_results1_MMMMPP}
\end{figure}

These results are further illustrated in Figure~\ref{f:sim_results1_MMMMPP}, which displays the distribution of population size estimates across the 50 simulated data sets. Overall, the results demonstrate that fitting the model on a substantially coarser discretisation than that used to generate the data can induce considerable negative bias in abundance estimation. However, once the fitted discretisation becomes sufficiently fine relative to the generating process, the remaining bias is small.

The computational cost increases rapidly with the size of the state space, from 1.04 seconds for $S = 49$ to 669 seconds for $S = 400$. This increase arises from repeated matrix computations involving matrix exponentials of size $S \times S$. The choice of discretisation is therefore a key practical consideration in the \emph{Move-SCR} framework, requiring a balance between computational feasibility and accurate approximation of the underlying continuous movement process. These results provide practical insight into how coarse discretisations may affect inference and how increasing state space resolution can improve estimation at substantial computational cost.

\end{document}